\documentclass[smallextended]{svjour3}       
\smartqed  
\usepackage{graphics,epsfig}
\usepackage{graphicx}
\usepackage{float}
\usepackage{amssymb,amstext,amsmath}
\usepackage{mathtools}
\usepackage{physics}
\usepackage{booktabs}
\usepackage[latin1]{inputenc}
\usepackage{epstopdf}
\newcommand{\bxi}{\boldsymbol{\xi}}
\newcommand{\llbracket}{[\![}
\newcommand{\rrbracket}{]\!]}
\journalname{Archive of Applied Mechanics}
\begin{document}

\title{Non-uniform plastic deformations of crystals undergoing anti-plane constrained shear\thanks{Y. Piao acknowledges financial support from China Scholarship Council (CSC).}
}

\titlerunning{Non-uniform plastic deformations at anti-plane constrained shear}        

\author{K. C. Le         \and
        Y. Piao 
}


\institute{K. C. Le (corresponding author)\at
              Lehrstuhl f\"{u}r Mechanik - Materialtheorie, Ruhr-Universit\"{a}t Bochum,
              \\D-44780 Bochum, Germany \\
              Tel.: +49-(0)234-3226033\\
              Fax: +49-(0)234-3206033\\
              \email{chau.le@rub.de}           
           \and
           Y. Piao \at
               Lehrstuhl f\"{u}r Mechanik - Materialtheorie, Ruhr-Universit\"{a}t Bochum,
              \\D-44780 Bochum, Germany
}

\date{Received: date / Accepted: date}

\maketitle

\begin{abstract}
The present paper studies non-uniform plastic deformations of crystals undergoing anti-plane constrained shear. The asymptotically exact energy density of crystals containing a moderately large density of excess dislocations is found by the averaging procedure. This energy density is extrapolated to the cases of extremely small or large dislocation densities. By incorporating the configurational temperature and the density of redundant dislocations, we develop the thermodynamic dislocation theory for non-uniform plastic deformations and use it to predict the stress-strain curves and the dislocation densities.

\end{abstract}

\section{Introduction}
Macroscopically observable plastic deformations of single crystals and polycrystalline materials are caused by nucleation, multiplication and motion of dislocations. When the number of dislocations increases, they may block each other leading to the work hardening \cite{Kuhlmann-Wilsdorf1989,Langer2010}. On the other side, as the dislocations move a large portion of plastic work is dissipated into heat causing different effects like the thermal softening \cite{Le2017a} or the formation of adiabatic shear bands \cite{Le2017b}. Therefore the understanding of irreversible thermodynamics of crystals containing dislocations is crucial in constructing the physically meaningful continuum theory of plasticity. After a long stagnation of the conventional phenomenological plasticity, the real progress has recently been made in the theory of dislocation mediated plastic flow proposed by Langer, Bouchbinder, and Lookman, called the LBL-theory for short (see \cite{Langer2010,Langer2015,Langer2016,Langer2017}). The breakthrough therein is to decouple the system of dislocated crystal into configurational and kinetic-vibrational subsystems. The configurational degrees of freedom describe the relatively slow, i.e. infrequent, atomic rearrangements that are associated with the irreversible movement of dislocations; the kinetic-vibrational degrees of freedom the fast vibrations of atoms in the lattice. The governing equations of LBL-theory are based on the kinetics of thermally activated dislocation depinning and irreversible thermodynamics of driven systems. This LBL-theory has been successfully used to simulate the stress-strain curves for copper over fifteen decades of strain rate, and for temperatures between room temperature and about one third of the melting temperature showing the excellent agreement with the experiments conducted by Follansbee and Kocks \cite{Follansbee1998}. The theory has been extended to include the interaction between two subsystems by Langer \cite{Langer2017} and used to simulate the stress-strain curves for aluminum and steel alloy \cite{Le2017a} which exhibit the thermal softening in agreement with the experiments conducted by Shi {\it et al.} \cite{Shi1997} and Abbot {\it et al.} \cite{Abbot2007}. It has been employed in predicting the formation of adiabatic shear band in steel HY-100 \cite{Le2017b} that shows the quantitative agreement with the experimental observations by Marchand and Duffy \cite{Marchand1988}.

The LBL-theory applies to the uniform plastic deformations, where the dislocations are redundant in the sense that their average Burgers vector vanishes. The extension of this theory to non-uniform plastic deformations including the excess dislocations \cite{Le2018} is based on the phenomenological free energy density proposed by Berdichevsky \cite{Berdichevsky2006}. Berdichevsky \cite{Berdichevsky2017} has shown later that the asymptotically exact free energy density of excess dislocations can be found. Based on the numerical simulation of excess dislocations in a twisted bar \cite{Weinberger2011}, he conjectured that the low-energy distribution of moderately large number of excess dislocations must be locally double-periodic. Taken this for granted, the two-scale homogenization technique (see, e.g., \cite{Bakhvalov1974,Bensoussan1978}) has been applied to derive the asymptotically exact formula for the energy density. However, as will be shown in this paper, the obtained formula without modification does not lead to the well-posed boundary-value problems within the continuum approach. Therefore, in order to use this formula in the thermodynamic dislocation theory we need to extrapolate it to extremely small and large excess dislocation densities. The aim of this paper is threefold. First, we extend Berdichevsky's formula for the energy density of excess screw dislocations to the anti-plane shear deformation. Then we provide the extrapolation of the obtained result to extremely small or large excess dislocation densities. Finally, we incorporate the formula for the energy density in the thermodynamic dislocation theory involving the redundant dislocations and configurational temperature and use it to compute the stress-strain curve and dislocation distribution in this problem. 
  
The paper is organized as follows. In Section 2 the averaging procedure for the ensemble of screw dislocations in crystals undergoing anti-plane constrained shear is developed. Section 3 provides the extrapolation of the energy density. In Section 4 the thermodynamic dislocation theory including this modified energy density of dislocated crystals is presented. Section 5 applies the proposed theory to the boundary-value problem of crystals undergoing anti-plane constrained shear. Section 6 shows the results of numerical simulations. Finally, Section 7 concludes the paper.

\section{Averaging procedure and energy of screw dislocations}
\begin{figure}[htb]
	\centering
	\includegraphics[width=6cm]{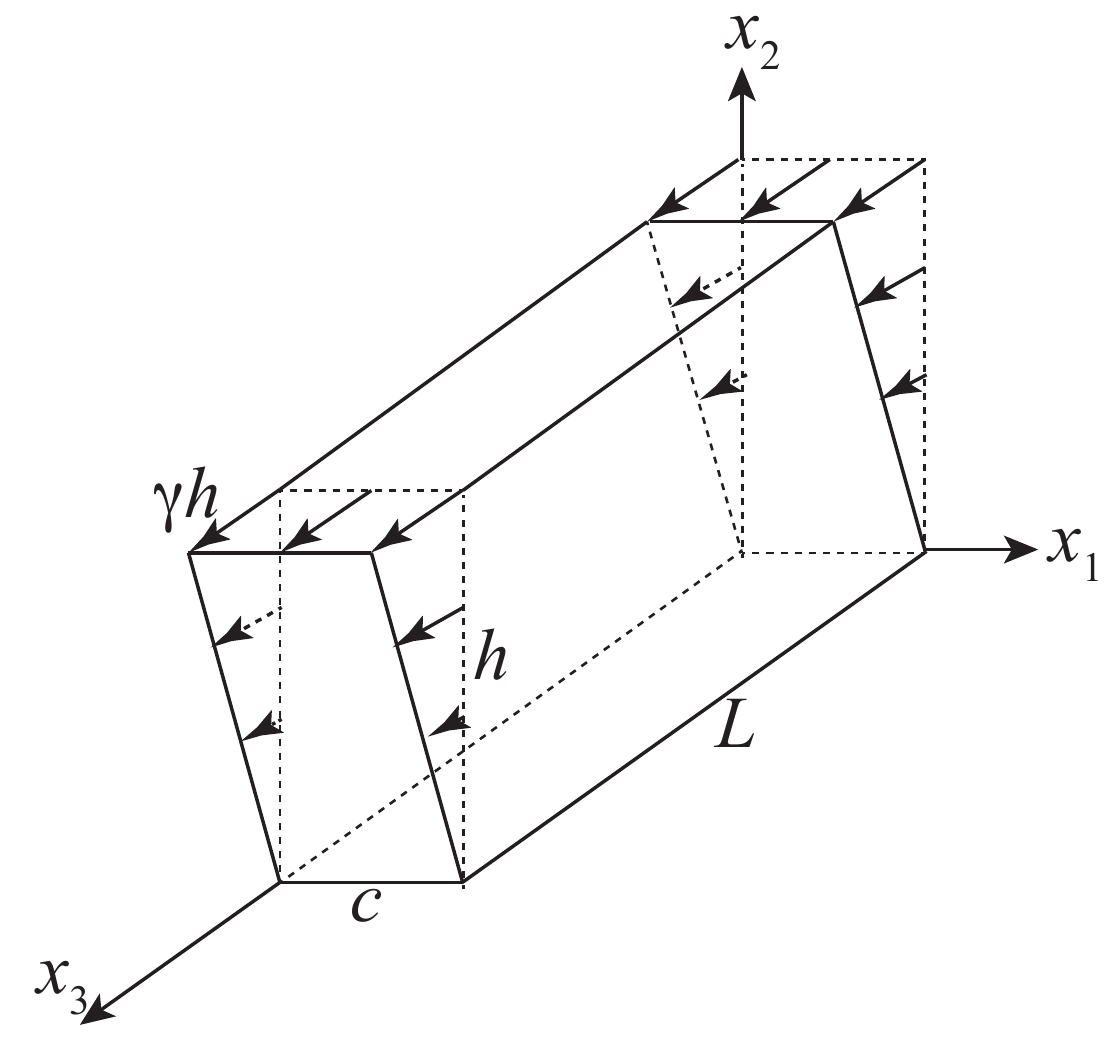}
	\caption{Anti-plane constrained shear}
	\label{fig:1}
\end{figure}
Consider a single crystal layer undergoing an anti-plane shear deformation. Let $A$ be the cross section of the layer perpendicular to the $x_{3}$-axis. For simplicity, we take $A$ as a rectangle, $A=(0,c)\times (0,h)$, with $c$ and $h$ being the width and the height of the cross section, respectively. We place this single crystal in a ``hard'' device with the prescribed displacement at the boundary $\partial A\times[0,L]$, with $L$ being the depth of the layer (see Fig.~\ref{fig:1})
\begin{equation}
\label{2.1}
w=\gamma (t) x_{2} \quad \text{at $\partial A \times [0,L]$}.
\end{equation}
Here $w(x_{1},x_{2},t)$ denotes the $x_3$-component of the displacement and $\gamma (t)$ corresponds to the overall shear regarded as a given function of time $t$. We assume that $c\ll h\ll L$. The problem is to predict the stress-strain curve as well as the dislocation density during the plastic deformation.

\subsection{A pair of screw dislocations:}

\begin{figure}[htb]
	\centering
	\includegraphics[width=7cm]{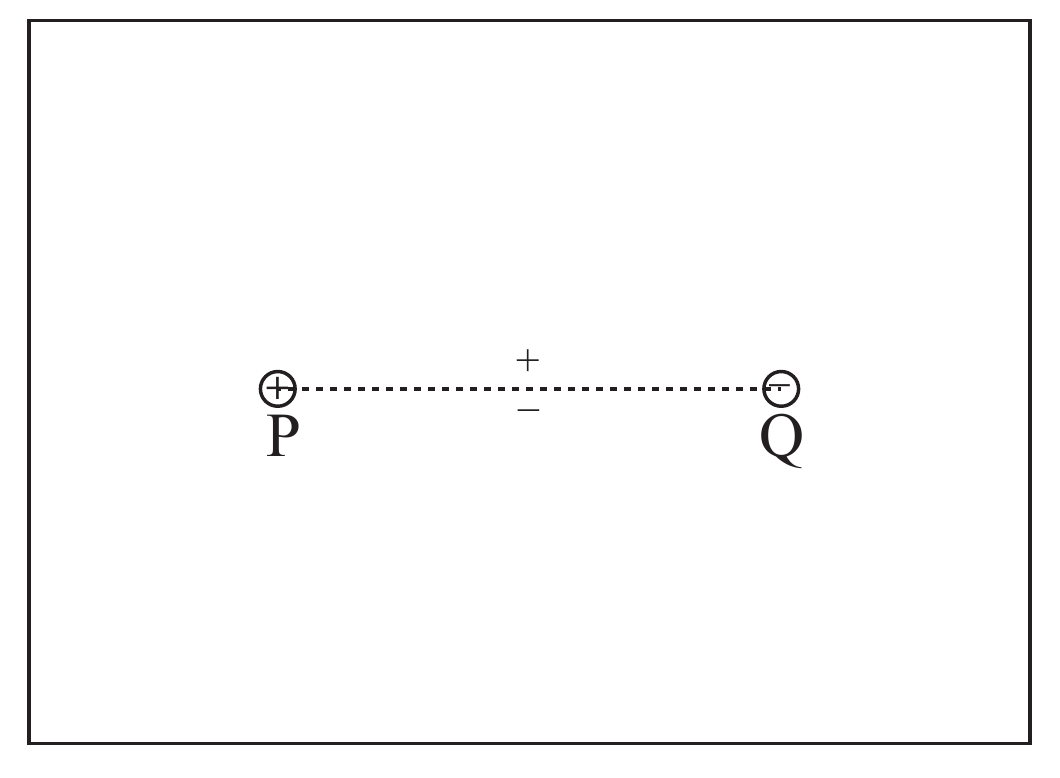}
	\caption{A cut creating a pair of dislocations}
	\label{fig:1a}
\end{figure}

Let us first consider the equilibrium with a fixed amount of shear $\gamma$. If $\gamma $ is large, then dislocations may occur in the equilibrium state of this crystal layer. Assume that a pair of screw dislocations is created by the well-known thought operations of cutting, shifting, and relaxing the crystal as shown schematically in Fig.~\ref{fig:1a}. Here a cut $\Lambda \times [0,L]$ is made along the straight dashed line $\Lambda $=PQ in the $(x_1,x_2)$-plane, and the atoms on the plus side of the cut is shifted in the $x_3$ direction through one lattice distance. Then the atoms are rejoined again and the whole crystal is relaxed. By these operations we have thus created a positive dislocation located at P and a negative dislocation located at Q, with the dislocation lines being parallel to the $x_3$-axis. Since the cut cannot reach the boundary $\partial A$ of the crystal's cross section due to the smooth displacement specified there, dislocations must always occur in pairs. In this sense the ``hard'' boundary conditions model the grain boundaries serving as obstacles and preventing dislocations to reach them. The displacement $w(\mathbf{x})$, with $\mathbf{x}=(x_{1},x_{2})$, in the relaxed equilibrium state suffers a jump on the line $\Lambda $ which is equal to the magnitude of Burgers' vector $b$, 
\begin{equation}\label{2.1a}
\llbracket w\rrbracket \equiv w^+-w^-=b \quad \text{on $\Lambda $},
\end{equation}
where $w^+$ and $w^-$ are the limiting values of $w$ on the upper and lower side of $\Lambda $, respectively. Gibbs variational principle states that the true displacement of the crystal in the relaxed equilibrium state minimizes the energy functional 
\begin{equation*}
I=\int_{A\setminus \Lambda } \frac{\mu}{2} ( w_{,1}^2+w_{,2}^2 ) \dd{x}
\end{equation*}
among all admissible displacements satisfying \eqref{2.1} and \eqref{2.1a}, where $\mu$ is the shear modulus and $\dd{x}=\dd{x}_1\dd{x}_2$. We get rid of the constraint \eqref{2.1a} and the cut by regarding function $w(\mathbf{x})$ as the distribution (or generalized function, \cite{Gelfand1964}). Then the derivatives of this generalized function are given by
\begin{equation*}
w_{,i}=bm_i \delta (\Lambda )+\tilde{w}_{,i},
\end{equation*}
where $m_i$ is the unit normal vector to $\Lambda $, $\delta (\Lambda )$ the Dirac delta function with the support $\Lambda $, and $\tilde{w}$ the multi-valued displacement defined on $A$. In what follows the Latin indices run from 1 to 2, and over repeated indices the summation is understood. We call $\beta _{3i}=bm_i \delta (\Lambda )$ the plastic distortion, while $\beta ^e_{3i}= \tilde{w}_{,i}$ the elastic distortion which is assumed to be regular everywhere except maybe at the dislocation line. Thus,
\begin{equation*}
w_{,i}=\beta _{3i}+\beta ^e_{3i}.
\end{equation*}
Since the total strain is equal to the elastic strain outside the cut, we remove the cut and reduce the above variational problem to the eigenstrain problem of minimizing the energy functional
\begin{equation*}
I=\int_{A} \frac{\mu}{2} [ (w_{,1}-\beta _{31})^2+(w_{,2}-\beta _{32})^2 ] \dd{x},
\end{equation*}
among all distributions satisfying \eqref{2.1} \cite{Le2010}. Changing the unknown function as $w=\gamma x_{2}+u(\mathbf{x})$, with $u=0$ at the boundary $\partial A$, we get the minimization problem 
\begin{equation*}
I=\int_A \frac{\mu}{2}\left[(u_{,1}-\beta _{31})^2 +( u_{,2}+\gamma -\beta _{32})^2 \right] \dd{x}\to \min_{u|_{\partial A}=0}.
\end{equation*}
The energy (per unit depth) of the crystal containing this pair of dislocations is defined as the minimum value of this functional, $\underline{I}$.

It is convenient to deal with the dual variational problem. The standard procedure (see \cite{Berdichevsky2009}) leads to minimizing the functional
\begin{equation}
\label{dual1}
\min_{\sigma_{3i}} \int_A [\sigma_{31}\beta_{31}+\sigma_{32}\beta_{32}-\sigma_{32}\gamma +\frac{1}{2\mu}(\sigma_{31}^2+\sigma_{32}^2 )] \dd{x}
\end{equation}
among all shear stresses $\sigma _{31}$ and $\sigma _{32}$ satisfying the equilibrium equation
\begin{equation*}
\sigma_{31,1}+ \sigma_{32,2}=0.
\end{equation*}
This equation is fulfilled if
\begin{equation*}
\sigma_{31}= \psi_{,2}, \quad \sigma_{32}=- \psi_{,1}.
\end{equation*}
Substituting these formulas into \eqref{dual1} and integrating the first two terms by parts, we obtain the dual minimization problem in terms of the stress function $\psi$,
\begin{equation}
\label{2.2}
J=\int_{A}\left[\frac{1}{2\mu}\left(\nabla\psi \right)^2+\alpha \psi +\psi_{,1}\gamma \right]\dd{x} \to \min_{\psi},
\end{equation}
where 
\begin{equation*}
\alpha =\beta _{32,1}-\beta _{31,2} =b[\delta (\mathbf{x}-\mathbf{x}^+)-\delta (\mathbf{x}-\mathbf{x}^-)].
\end{equation*}
Note that the energy of crystal containing these dislocations, $\underline{I}$, equals to the minimum of $J$ taken with minus sign, $\underline{I}=-\underline{J}$. It turns out that, if the jump of $w$ is constant on $\Lambda $, the energy is infinite. Therefore this variational problem needs a regularization.

\begin{figure}[htb]
	\centering
	\includegraphics[width=7cm]{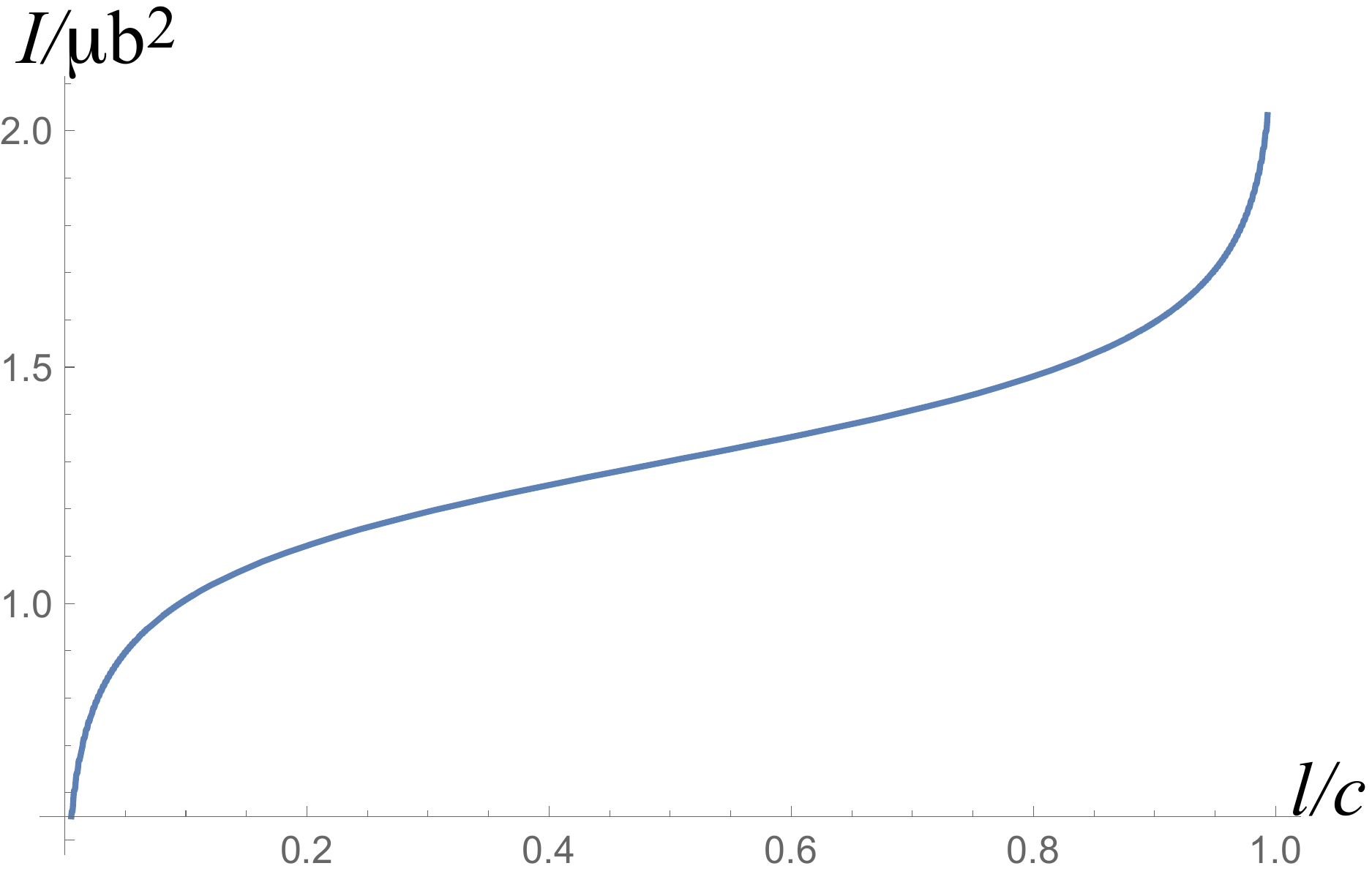}
	\caption{Energy of dislocation dipole as function of $l/c$ ($\gamma=0$)}
	\label{fig:dipole0}
\end{figure}

\begin{figure}[htb]
	\centering
	\includegraphics[width=7cm]{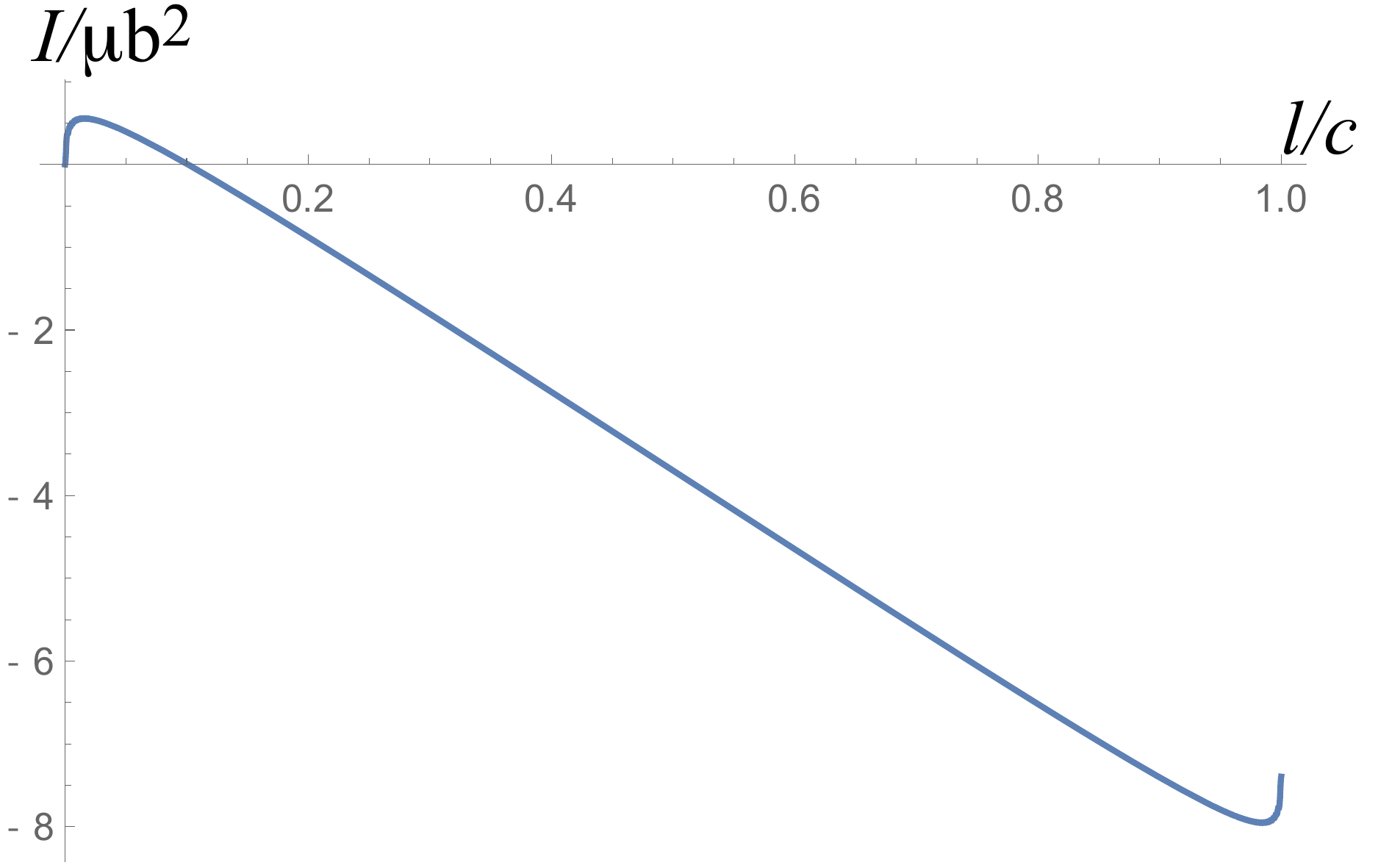}
	\caption{Energy of dislocation dipole (with the constant term $\frac{1}{2}\mu \gamma^2 ch$ being removed) as function of $l/c$ ($\gamma=0.001$)}
	\label{fig:dipole1}
\end{figure}

The simplest regularization of the above variational problem is to use in \eqref{2.2} the regularized dislocation density $\alpha_r=b[\delta_{r_{0}}(\mathbf{x}-\mathbf{x}^+)-\delta _{r_{0}}(\mathbf{x}-\mathbf{x}^-)]$ instead of $\alpha $, where
\begin{equation*}
\delta_{r_{0}}(\mathbf{x}-\bxi)=
\begin{cases}
\frac{1}{\pi r_{0}^2} & |\mathbf{x}-\bxi |<r_{0},\\
0 & \text{otherwise.}
\end{cases}
\end{equation*}
Here, $r_0$ is the radius of a small circle with the center at $\bxi$, interpreted as the dislocation core.\footnote{Other regularizations are also possible (see, e.g., \cite{Cai2006,Aifantis2009,Po2014}).} Varying the energy functional \eqref{2.2}, with $\alpha $ being replaced by $\alpha _r$, we derive the following boundary-value problem
\begin{equation}
\label{2.3}
  \begin{cases}
  \nabla^2\psi=\mu \alpha_r & \text{in $A$}, \\
 \psi_{,1}=-\mu\gamma & \text{on $\partial A_{1}=(0,x_{2})$ and $\partial A_{3}=(c,x_{2})$}, \\
 \psi_{,2}=0 & \text{on $\partial A_{2}=(x_{1},0)$ and $\partial A_{4}=(x_{1},h)$}.
  \end{cases}
\end{equation}
The boundary conditions in \eqref{2.3} will be simplified if we change the unknown function as follows: $\psi=-\mu\gamma x_{1}+\varphi$. Then, in terms of $\varphi $, the variational problem becomes
\begin{equation}
\label{2.4}
J=\int_{A}\left[-\frac{1}{2}\mu \gamma ^2+\frac{1}{2\mu}\left(\nabla\varphi \right)^2-\mu \gamma x_1\alpha _r+\alpha_r\varphi \right]\dd{x} \to \min_{\varphi}.
\end{equation}
This variational problem implies the Poisson equation subjected to the Neumann boundary condition for $\varphi$  
\begin{equation*}
  \begin{cases}
  \nabla^2\varphi=\mu \alpha _r & \text{in $A$}, \\
{\varphi }_{,n}=0 & \text{on $\partial A$}, 
  \end{cases}
\end{equation*}
where ${\varphi }_{,n}$ is the derivative in the normal direction to the boundary of $A$. Inserting the minimizer $\check{\varphi}$ into the functional $J$ and making use of Clapeyron's theorem, we get for the energy of crystal containing two dislocations
\begin{equation*}
\underline{I}=-\underline{J}=\int_{A}(\frac{\mu}{2}\gamma^2+\mu \gamma x_1\alpha _r-\frac{1}{2}\alpha_r\check{\varphi })\dd{x}.
\end{equation*}
As an example, let us compute this energy in the case, when a positive dislocation is located at $(c/2-l/2,h/2)$ and a negative one at $(c/2+l/2,h/2)$. The dimensionless energy $\underline{I}/\mu b^2$ is shown in Fig.~\ref{fig:dipole0} (for $\gamma =0$) and Fig.~\ref{fig:dipole1} (for $\gamma =0.001$, with the constant term $\frac{1}{2}\mu \gamma^2 ch$ being removed) as function of the distance $l/c$. In both cases the energy has a local minimum at $l=0$. Thus, if there is no thermal fluctuation, then the nucleation of dislocation dipole is energetically not preferable. However, as shown in \cite{Berdichevsky-Le2002}, the presence of thermal fluctuation changes the situation. Now, for each temperature there will be a certain density of the dislocation dipoles, with quite small mean distance between dislocations in the dipole. If the external field is applied, then the dipole can even be dissolved into freely moving dislocation if the energy barrier can be overcome. Note that the larger is the field, the smaller the energy barrier, so with the thermal fluctuation it becomes easier to dissolve the dipoles if the applied shear stress is large enough.

\subsection{A large number of screw dislocations:}
Now we allow an equal large number of screw dislocations of opposite signs to enter the crystal layer simultaneously. Due to the almost translational invariance in the $x_2$-direction, we assume that the positive and negative dislocations are well-separated and distributed symmetrically about the straight line $x_1=c/2$. In this case $\beta _{31}=0$, while 
\begin{equation}
\label{beta}
\beta_{32}(\mathbf{x})\equiv \beta(\mathbf{x})=\sum^N_{i=1}b\delta(\Lambda_i),
\end{equation}
with $\Lambda _i$ being the straight segments parallel to the $x_1$-axis with the middle points lying on the line $x_1=c/2$. The regularized dislocation density is given by
\begin{equation*}
\alpha_r(\mathbf{x})=\sum^N_{i=1}b\delta_{r_{0}}(\mathbf{x}-\mathbf{x}^+_i)-\sum^N_{i=1}b\delta_{r_{0}}(\mathbf{x}-\mathbf{x}^-_i),
\end{equation*}
where $\mathbf{x}^+_i$ and $\mathbf{x}^-_i$ are the positions of the positive and the negative dislocations, respectively (the end-points of $\Lambda _i$). Thus, the dislocation density is a piecewise constant and fast changing function of the coordinates. We further assume that function $\alpha_r(\mathbf{x})$ is locally double-periodic, with the characteristic period being much smaller than $c$. Following Berdichevsky \cite{Berdichevsky2017}, we split the regularized dislocation density into the average dislocation density denoted by $\bar{\alpha}_r$ and the fluctuation denoted by $\alpha^\prime _r$
\begin{equation}
\label{2.7}
\alpha_r=\bar{\alpha}_r+\alpha^\prime _r.
\end{equation}
Here the averaging over the cell is defined as
\begin{equation*}
\bar{\alpha}_r=\frac{1}{|C_x|}\int_{C_x}\alpha _r \dd{x},
\end{equation*}
where $C_x$ is the periodic cell in the $(x_1,x_2)$-plane, while $|C_x|$ denotes its area. Thus, $\bar{\alpha}_r$ is a slowly changing function of the coordinates. We call $\rho ^g=|\bar{\alpha}_r|/b$ the density of excess dislocations (or average dislocation density). This decomposition gives rise to the decomposition of the stress function $\varphi $ and the plastic slip $\beta$ as well
\begin{equation}
\label{2.8}
\begin{split}
\varphi&=\bar{\varphi}+\varphi^\prime ,\\
\beta&=\bar{\beta}+\beta^\prime .
\end{split}
\end{equation}
Note that function $\beta $ defined in \eqref{beta} is non-periodic and equals the sum of generalized functions concentrated on the cut lines $\Lambda _i$. Therefore, the integral over $C_x$ of $\beta $ equals the sum of line integrals over those segments $\Lambda _i$ lying within this cell. It is easy to see that (cf. \cite{Nye1953})
\begin{equation*}
\bar{\alpha}_r=\bar{\beta}_{,1}.
\end{equation*}
Inserting the decomposed dislocation density \eqref{2.7} and the decomposed stress function \eqref{2.8}$_1$ into the energy functional \eqref{2.4}, we get $J=J_1+J_2$, where
\begin{equation*}
J_1=\int_{A} \left[ -\frac{1}{2}\mu \gamma ^2-\mu \gamma x_1 \bar{\alpha}_r +\frac{1}{2\mu}\left(\nabla \bar{\varphi }\right)^2+\bar{\alpha }_r
\bar{\varphi }\right] \dd{x} , 
\end{equation*}
while
\begin{multline}
\label{2.11}
J_2=\int_{A}(-\mu \gamma x_1 \alpha^\prime _r+\alpha ^\prime _r \bar{\varphi }) \dd{x} +\int_A \left( \frac{1}{\mu }\nabla \bar{\varphi }\cdot \nabla \varphi ^\prime +\bar{\alpha }_r
\varphi ^\prime \right) \dd{x} 
\\
+\int_{A} \left[ \frac{1}{2\mu}\left(\nabla \varphi ^\prime \right)^2+\alpha ^\prime _r
\varphi ^\prime \right] \dd{x}.
\end{multline}

Based on this decomposition the minimization of $J$ splits into the minimization of $J_1$ among $\bar{\varphi}$ and then $J_2$ among $\varphi ^\prime $, provided $\bar{\varphi}$ is known. It is easy to show that the negative minimum value of $J_1$ coincides with the energy of average elastic strain 
\begin{equation}
\label{2.12}
-\underline{J}_1=\int_A \frac{1}{2}\mu (\gamma -\bar{\beta})^2 \dd{x}.
\end{equation}
Concerning the functional $J_{2}$ we see that, due to the Euler equation for $\bar{\varphi }$, the second integral in \eqref{2.11} vanishes. The first integral is small and can be neglected. The minimization of the last integral in \eqref{2.11} among periodic functions $\varphi ^\prime $ for the hexagonal periodic dislocation structure has been solved by Berdichevsky \cite{Berdichevsky2017}. The combination of his result with \eqref{2.12} leads to the following statement: the energy density of crystal containing excess dislocations equals the sum of energy density of macroscopic elastic strain and energy density of excess dislocations $\phi_m(\rho ^g)$
\begin{equation*}
\phi =\frac{\mu}{2}\left(\gamma-\bar{\beta}\right)^2+\phi _{m}(\rho ^g),
\end{equation*}
where
\begin{equation}
\label{2.last}
\phi _{m}(\rho ^g)=\mu b^2 \rho ^g\left[ \phi^*+\frac{1}{4\pi}\ln \frac{1}{b^2\rho ^g} \right].
\end{equation}
Here $\phi^*$ is a parameter depending on the periodic dislocation structure. For the hexagonal periodic dislocation structure $\phi^*=-0.105$.

\section{Extrapolation of energy density of excess dislocations}
\begin{figure}[htb]
	\centering
	\includegraphics[width=7cm]{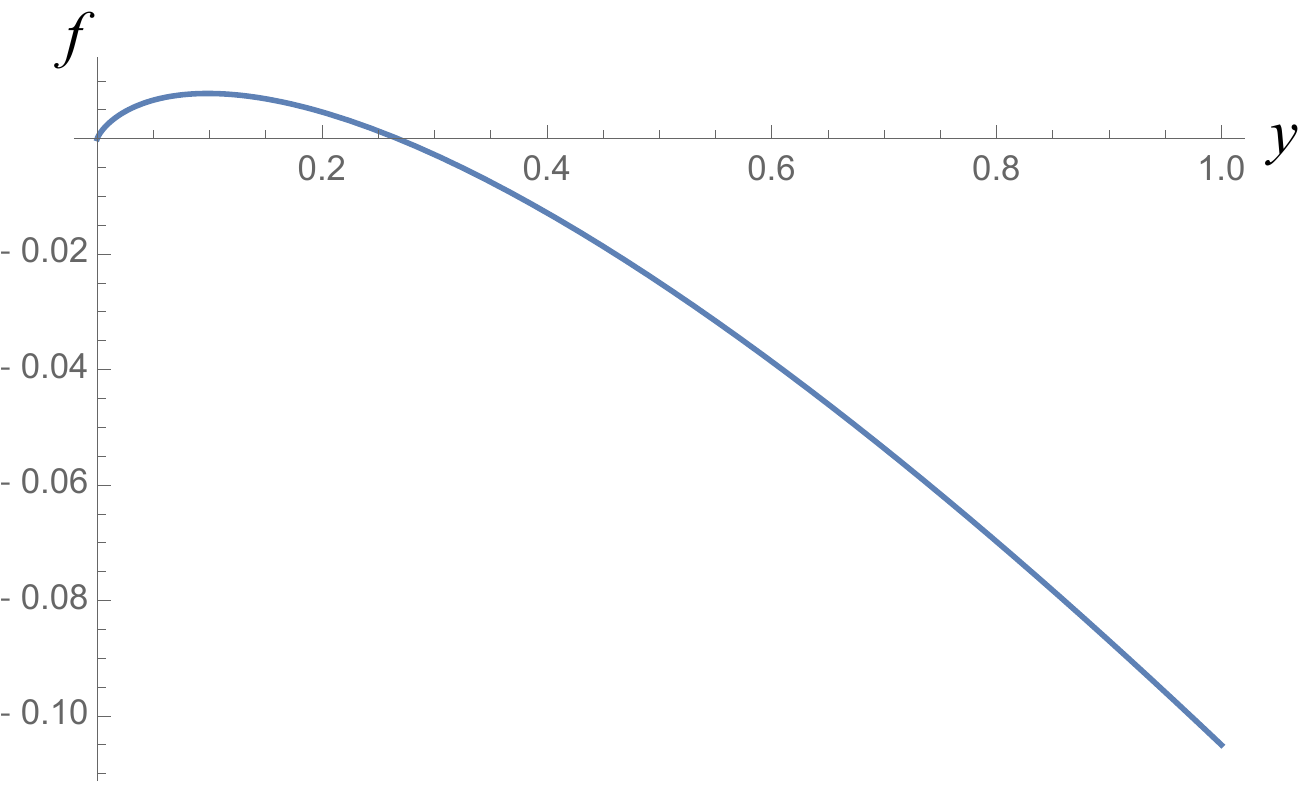}
	\caption{Dimensionless energy density $f=\phi _{m}/\mu$ versus dimensionless dislocation density $y=b^2\rho ^g$}
	\label{fig:Edensity}
\end{figure}

According to \eqref{2.last} the dimensionless energy density of excess dislocations  can be written as 
\begin{equation}
\label{3.1}
f(y)\equiv \phi _{m}/\mu =y(\phi^*-\frac{1}{4\pi}\ln y),
\end{equation}
where $y=b^2\rho ^g$ is the dimensionless dislocation density. The plot of this function for $y\in (0,1)$ is shown in Fig.~\ref{fig:Edensity}.
Function \eqref{3.1} possesses three remarkable properties. First, $f^\prime (0)=\infty $. Second, $f(y)$ is concave. Third, $f(y)$ tends to $-\infty$ when $y\to \infty$. These properties make the application of \eqref{3.1} to the determination of average plastic slip via energy minimization within the continuum approach problematic. For, the well-posedness of the boundary value problems within the continuum approach requires the convexity of the energy density and the regularity of its derivative with respect to $\rho^g$ (the latter is needed for the regularity of the back-stress). When looking closer at the assumptions made in deriving formula \eqref{2.last}, we see that these assumptions may be violated for the extremely small or large dislocation densities. Such extreme values of dislocation densities may occur near the head and the tail of the dislocation pile-up. Therefore the energy density \eqref{2.last} needs be extrapolated to these extremely small or large dislocation densities. 

We propose the following extrapolation for the free energy density
\begin{equation}
\label{3.2}
\phi _{m}(\rho ^g)=\mu b^2 \rho ^g \left(\phi^*+\frac{1}{4\pi}\ln{\frac{1}{k_0+b^2\rho ^g}}\right)+\frac{1}{8\pi }\mu k_1 (b^2\rho ^g)^2.
\end{equation}
with $k_0$ and $k_1$ being two new material constants. The small constant $k_0$ corrects the behavior of the derivative of energy at $\rho ^g=0$, while the last term containing $k_1$ corrects the behavior of the energy at large density of the excess dislocations. We choose $k_0$ and $k_1$ so that: (i) the energy density is close to the asymptotic exact energy density for moderate dislocation densities, (ii) $\phi _{m}(\rho ^g)$ is the convex function for all positive $\rho ^g$. The latter requirement guarantees the existence of the energy minimizer. To investigate the convexity we compute the second derivative of 
\begin{equation}
\label{3.3}
f(y)\equiv \phi_m/\mu =y \left(\phi^*+\frac{1}{4\pi}\ln{\frac{1}{k_0+y}}\right)+\frac{1}{8\pi } k_1 y^2
\end{equation}
as function of $y=b^2\rho ^g$. The simple calculation shows that
\begin{equation*}
\frac{d^2f}{dy^2}=\frac{k_1y^2+(2k_0k_1-1)y+k_1k_0^2-2k_0}{4\pi (k_0+y)^2}.
\end{equation*}
For function $\phi_m$ to be convex the numerator must be positive for $y>0$. Since the roots of this quadratic function are
\begin{equation*}
y_{1,2}=\frac{1-2k_0k_1\pm \sqrt{1+4k_0k_1}}{2k_1},
\end{equation*}
it is sufficient to require the largest root to be negative. This gives the following constraint for the coefficients $k_0$ and $k_1$
\begin{displaymath}
k_0k_1>2.
\end{displaymath}

\begin{figure}[htb]
	\centering
	\includegraphics[width=13cm]{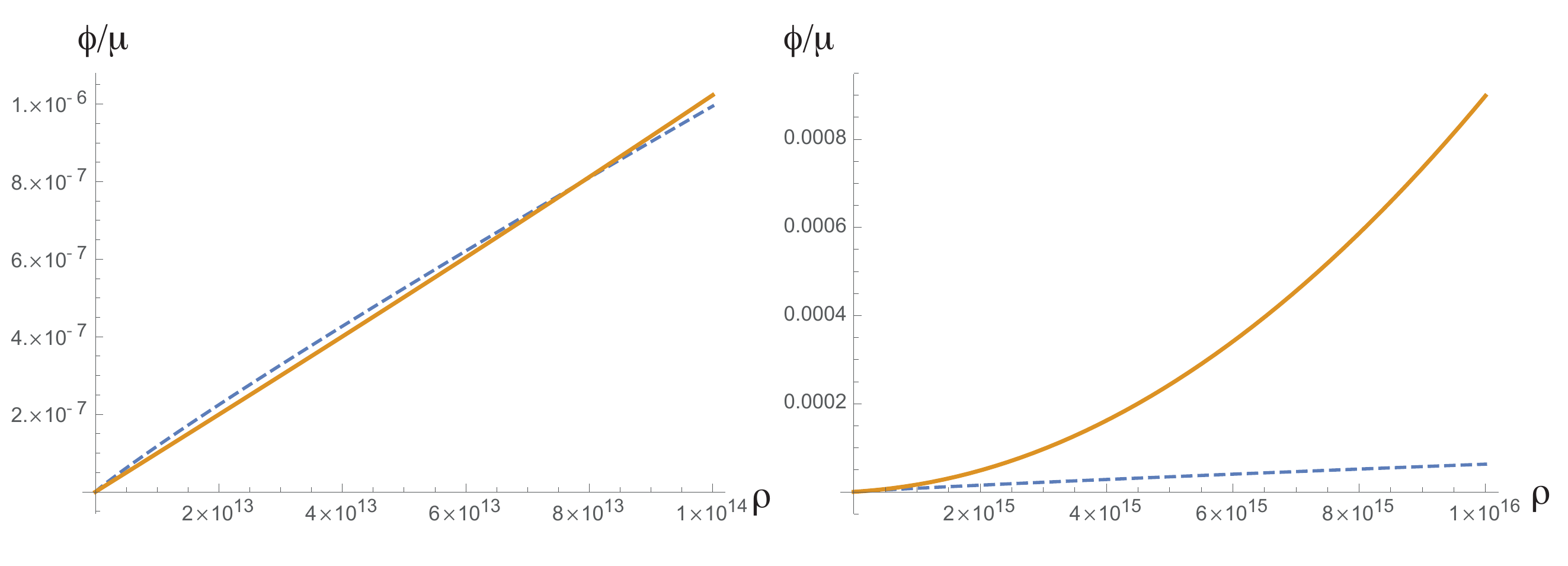}
	\caption{The dimensionless energy density $\phi _{m}/\mu$ versus the density of excess dislocations $\rho ^g$: (i) \eqref{2.last} (dashed line), (ii) \eqref{3.2} (bold line)}
	\label{fig:EnergyModified}
\end{figure}

Fig.~\ref{fig:EnergyModified} shows the comparison between two dimensionless energy densities $\phi_m/\mu $ defined in accordance with \eqref{2.last} and \eqref{3.2} within the range $\rho^g\in (0,10^{14}$/m$^2$) (on the left) and the range $\rho^g\in (0,10^{16}$/m$^2$) (on the right). Here we choose $b=10^{-10}$m, $k_0=10^{-6}$, $k_1=2.1\times 10^{6}$. We see that the two energy densities are nearly the same in the range $\rho^g\in (0,10^{14}$/m$^2$), but differ essentially for $\rho^g$ larger than $10^{14}$/m$^2$.

\section{Thermodynamic dislocation theory}
\label{sec:4}

Now let us incorporate the formula for the energy density in the thermodynamic dislocation theory involving the redundant dislocations and configurational temperature proposed recently by Le \cite{Le2018}. As discussed in Section 2, the redundant dislocations, nucleated by the thermal fluctuation, exist in form of dislocation dipoles, whose mean distance is much smaller than the size of the periodic cell of excess dislocations. By this reason the dislocation dipoles do not affect the average dislocation density which is identified with the density of excess dislocations. Following Kr\"oner \cite{Kroener1992} and Langer \cite{Langer2016} we require that the free energy density depends on the average elastic shear strain $\gamma-\bar{\beta}$, the densities of redundant dislocations $\rho ^r$ and excess dislocations $\rho ^g=|\bar{\beta}_{,1}|/b$, the kinetic-vibrational temperature $T$, and the configurational temperature $\chi $. We restrict ourself to the isothermal processes, so the kinetic-vibrational temperature $T$ is assumed to be constant and can be dropped in the list of arguments of the free energy density. Our main assumption for the free energy density is
\begin{equation}\label{eq:4.1}
\phi = \frac{1}{2}\mu (\gamma-\bar{\beta})^2+e_D\rho ^r+\phi _{m}(\rho^g) -\chi (- \rho \ln (a^2 \rho )+\rho ),
\end{equation}
where $\rho=\rho^r+\rho^g$ is the total density of dislocations and $a$ the mean distance between dislocations in the saturated state. The first term in \eqref{eq:4.1} describe the energy density of crystal due to the macroscopic elastic strain. The second term is the energy density of redundant dislocations, with $e_D$ being the energy of the dislocation dipole. The third term is the energy density of excess dislocations, where $\phi _{m}(\rho^g)$ is the extrapolated energy density taken in accordance with \eqref{3.2}. The last term has been introduced by Langer \cite{Langer2016}, with $S_C=-\rho \ln (a^2 \rho )+\rho $ being the configurational entropy of dislocations.

With this free energy density we can now write down the energy functional of the crystal
\begin{equation*}
I[\bar{\beta }(\mathbf{x},t),\rho ^r(\mathbf{x},t),\chi (\mathbf{x},t))]=\int_{A} \phi (\bar{\varepsilon}^e,\rho ^r, \rho ^g,\chi )\dd{x}.
\end{equation*}
Note that the average plastic slip satisfies the kinematic boundary condition 
\begin{equation}
\label{eq:4.3a}
\bar{\beta}(0,x_2,t)=\bar{\beta}(c,x_2,t)=0.
\end{equation}

Under the increasing overall shear strain $\gamma (t)$ the shear stress also increases, and when it reaches the Taylor stress, dislocation dipoles dissolve into freely moving dislocations. The latter dislocations move under the action of shear stress until they are trapped again by dislocations of opposite sign. During this motion dislocations always experience the resistance causing the energy dissipation. The increase of dislocation density as well as the increase of configurational temperature also lead to the energy dissipation. Neglecting the dissipation due to the internal viscosity associated with the strain rate, we propose the dissipation potential in the form 
\begin{equation}\label{eq:4.4}
D(\dot{\bar{\beta}},\dot{\rho},\dot{\chi})=\tau _Y |\dot{\bar{\beta }}|+\frac{1}{2}d_\rho \dot{\rho }^2+\frac{1}{2}d_\chi \dot{\chi }^2,
\end{equation}
where $\tau_Y$ is the flow stress during plastic yielding, $d_\rho $ and $d_\chi $ are still unknown functions, to be determined later. The first term in \eqref{eq:4.4} is the plastic power which is assumed to be a homogeneous function of first order with respect to the plastic slip rate \cite{Puglisi2005}. The other two terms describe the dissipation caused by the multiplication of dislocations and the increase of configurational temperature \cite{Langer2010}. Based on Hooke's law, Orowan's equation and the kinetics of dislocation depinning \cite{Langer2010}, the following equation holds true for $\tau _Y$
\begin{equation}
\label{eq:4.5}
\dot{\tau }_Y=\mu \frac{q_0}{t_0}\left[ 1-\frac{q(\tau_Y,\rho)}{q_0}\right] .
\end{equation}
In Eq.~\eqref{eq:4.5} $q_0/t_0=\dot{\gamma }$ is the rate of the shear strain, assumed here to be positive, with $t_0=10^{-12}$s being the characteristic microscopic time scale. The rate of plastic slip is $\dot{\bar{\beta}}=q(\tau_Y,\rho)/t_0$, where
\begin{equation*}
q(\tau_Y,\rho)=b\sqrt{\rho }\exp \left[ -\frac{1}{\theta }e^{-\tau _Y/\tau _T}\right] .
\end{equation*}
In this equation the dimensionless temperature is introduced as $\theta =T/T_P$, with $T_P$ being the pinning energy barrier, and $\tau _T=\mu _T b\sqrt{\rho }$ the Taylor stress. Since the dislocation mediated plastic flow is the irreversible process, we derive the governing equations from the following variational principle: the true average plastic slip $\check{\bar{\beta }}(\mathbf{x},t)$, the true density of redundant dislocations $\check{\rho }^r(\mathbf{x},t)$, and the true configurational temperature $\check{\chi }(\mathbf{x},t)$ obey the variational equation
\begin{equation}
\label{eq:4.8}
\delta I+\int_{\mathcal{A}} \left( \frac{\partial D}{\partial \dot{\bar{\beta }}}\delta \bar{\beta }+\frac{\partial D}{\partial \dot{\rho }}\delta \rho +\frac{\partial D}{\partial \dot{\chi }}\delta \chi \, \right) \dd{x}=0
\end{equation}
for all variations of admissible fields $\bar{\beta }(\mathbf{x},t)$, $\rho ^r(\mathbf{x},t)$, and $\chi (\mathbf{x},t)$ satisfying the constraints \eqref{eq:4.3a}. 

Taking the variation of $I$ with respect to three unknown functions $\bar{\beta }$, $\rho ^r$, and $\chi $ and requiring that Eq.~\eqref{eq:4.8} is satisfied for their admissible variations, we get three equations
\begin{equation}
\label{eq:4.10}
\begin{split}
\tau + \frac{1}{b}(\varsigma \, \text{sign} \bar{\beta}_{,1})_{,1}-\tau _Y+\frac{1}{b}(d_\rho \dot{\rho } \, \text{sign} \bar{\beta}_{,1})_{,1}=0, 
\\
e_D+\chi \ln (a^2\rho) +d_\rho \dot{\rho }=0, 
\\
\rho \ln (a^2\rho) -\rho+d_\chi \dot{\chi }=0,
\end{split}
\end{equation}
where $\tau =\mu (\gamma -\bar{\beta})$ is the shear stress, while $\varsigma=\partial\psi /\partial \rho ^g$. The first equation of \eqref{eq:4.10}, valid under the condition $\dot{\beta}>0$, can be interpreted as the balance of microforces acting on dislocations. This equation is subjected to the Dirichlet boundary condition \eqref{eq:4.3a}.  

We require that, for the uniform total and plastic deformations, system \eqref{eq:4.10}  reduces to the system of equations of LBL-theory \cite{Langer2010}
\begin{align}
\dot{\tau } & =  \mu \frac{q_0}{t_0}\left[1-\frac{q(\tau ,\rho )}{q_0}\right] ,  \notag
\\
\dot{\chi } & =  \mathcal{K} \tau \frac{q(\tau ,\rho )}{t_0} \left[ 1-\frac{\chi }{\chi ^{ss}(q)} \right], \label{eq:4.11}
\\
\dot{\rho } & = \mathcal{K}_\rho \frac{\tau }{a^2\nu (T,\rho ,q_0)^2}\frac{q(\tau ,\rho )}{t_0}\left[ 1-\frac{\rho }{\rho ^{ss}(\chi )} \right] . \notag
\end{align}
As compared to the original equations derived in \cite{Langer2010} there are some changes in notations to make them consistent with those employed in this paper: the shear stress is denoted by $\tau $ instead of $\sigma $, the shear strain rate by $\dot{\gamma }$ instead of $\dot{\epsilon }$, while the plastic slip rate by $\dot{\bar{\beta }}$ instead of $\dot{\varepsilon}^p$. The steady-state configurational temperature is denote by $\chi ^{ss}$, while the steady-state dislocation density is
\begin{displaymath}
\rho ^{ss}(\chi )=\frac{1}{a^2}e^{-e_D/\chi }.
\end{displaymath}
Finally, $\nu (T,\rho ,q_0)$ is defined as follows
\begin{displaymath}
\nu (T,\rho ,q_0)=\ln \left( \frac{T_P}{T}\right) - \ln \left[ \frac{1}{2}\ln \left( \frac{b^2\rho }{q_0^2}\right) \right] .
\end{displaymath}
Since the total and plastic deformation are uniform, the second and fourth terms in \eqref{eq:4.10}$_1$ disappear, so $\tau =\tau _Y$, and in combination with Eq.~\eqref{eq:4.5}, this leads to \eqref{eq:4.11}$_1$. Two remaining equations of \eqref{eq:4.10}$_{2,3}$ reduce to \eqref{eq:4.11}$_{2,3}$  if we choose
\begin{align}
\label{eq:4.12}
d_\chi &=\frac{\rho -\rho \ln (a^2\rho) }{\mathcal{K} \tau_Y \frac{q(\tau_Y ,\rho )}{t_0} \left[ 1-\frac{\chi }{\chi ^{ss}(q)} \right] },
\\
d_\rho &=\frac{-e_D-\chi \ln (a^2\rho)}{\mathcal{K}_\rho \frac{\tau_Y }{a^2\nu (T,\rho ,q_0)^2}\frac{q(\tau_Y ,\rho )}{t_0}\left[ 1-\frac{\rho }{\rho ^{ss}(\chi )} \right]} . \label{eq:4.13}
\end{align}
Note that, for $\rho $ changing between 0 and $\rho ^{ss}<1/a^2$, both numerators on the right-hand sides of \eqref{eq:4.12} and \eqref{eq:4.13} are positive, and the dissipative potential \eqref{eq:4.4} is positive definite as required by the second law of thermodynamics.

\section{Anti-plane shear deformation}
We turn back to the single crystal layer undergoing an anti-plane shear deformation with the increasing overall shear strain $\gamma (t)$ such that $\dot{\gamma}$=const. We aim at determining the average plastic slip, the densities of total and excess screw dislocations, the configurational temperature, and the stress-strain curve as function of $\gamma (t)$ by the thermodynamic  dislocation theory proposed in the previous Section.
 
When the applied shear stress exceeds the Taylor's stress, dislocation dipoles dissolve into freely moving dislocations. This applied stress drive the positive dislocations to the left and the negative ones to the right. After a short time these free dislocations will either be trapped by the dislocations of opposite sign or be blocked near the grain boundaries acting as the obstacles. Thus, the dislocations of the same sign piling up against the left and right boundaries become excess dislocations occupying the boundary layers. Since the thickness of the boundary layers is quite small compared to the width of the crystal, $c$, the average plastic slip $\bar{\beta }$ is nearly uniform in the middle of the specimen. Neglecting a small non-uniformity of $\bar{\beta }$ in the boundary layers, we reduce the determination of $\tau_Y$, $\rho $, and $\chi $ in the first approximation to the solution of \eqref{eq:4.11}, with $\tau $ being replaced by $\tau _Y$ and $q_0=t_0 \dot{\gamma }$. After knowing $\tau _Y$, $\rho $, and $\chi $, the variational equation \eqref{eq:4.8} reduces to minimizing the following ``relaxed'' energy functional
\begin{equation*}
I_d=h L\int^c_0\left[ \frac{1}{2}\mu\left(\gamma-\bar{\beta}\right)^2+\mu f(b|\bar{\beta}_{,1}|) + \tau_Y(\gamma) \bar{\beta } \right] \dd{x}_1
\end{equation*}
among $\bar{\beta }$ satisfying \eqref{eq:4.3a}, provided the sign of $\dot{\bar{\beta}}$ is positive during the loading course. Here we take into account that, due to the almost translational invariance in the $x_2$-direction, $\bar{\beta }$ depends only on $x_1$.
  
It is convenient to introduce the following dimensionless coordinates and quantities
\begin{equation*}
\bar{I}_d=\frac{I_d}{\mu b L h}, \quad \bar{x}=\frac{x_1}{b}, \quad \bar{c}=\frac{c}{b}, \quad g(\gamma)=\frac{\tau _Y(\gamma)}{\mu },
\end{equation*}
in terms of which the above functional becomes
\begin{equation*}
I_d=\int^c_0\left[ \frac{1}{2}\left(\gamma-\beta\right)^2+ f(|\beta ^\prime |) + g(\gamma) \beta \right] \dd{x}
\end{equation*}
where the prime denotes the derivative of a function with respect to its argument and, since we shall deal only with the dimensionless quantities, the bar over them will be omitted for short. Up to an unessential constant this functional can be reduced to
\begin{equation}
\label{5.2}
I_d=\int^c_0\left[ \frac{1}{2}\left(\gamma_l-\beta\right)^2+ f(|\beta ^\prime |) \right] \dd{x},
\end{equation}
where $\gamma_l(\gamma)=\gamma-g(\gamma)$. 

Following the method of solution of the dislocation pile-up problem developed in \cite{Berdichevsky-Le07}, we look for the minimizer in the form
\begin{equation*}  
\beta(x)=
  \begin{cases}
    \beta_1(x) & \text{for $x \in (0,l)$}, \\
    \beta_m & \text{for $x \in (l,c-l)$}, \\
    \beta_1(c-x) & \text{for $x \in (c-l,c)$}, \\
  \end{cases}
\end{equation*}
where $\beta _1(x)$ is an unknown increasing function, $\beta_m$ is a constant, $l$ an unknown length, $0\le l\le c/2$, and $\beta_1(l)= \beta_m$ at $x=l$. With this Ansatz, the functional becomes
\begin{equation}
\label{5.3}
I_d=2\int^{l}_0\left[\frac{1}{2}(\gamma_l-\beta_1)^2+f(\beta ^\prime_1)\right]\dd{x}
+\frac{1}{2}(\gamma_l-\beta_m)^2(c-2l).
\end{equation} 
Varying this energy functional with respect to $\beta_{1}$ we get the equation 
\begin{equation}
\label{5.4}
f^{\prime \prime }(\beta _1^\prime )\beta _{1}^{\prime \prime }+\beta _1= \gamma_l(\gamma),
\end{equation}
which is subjected to the boundary conditions
\begin{equation*}
\beta_1(0)=0,\quad \beta_1(l)=\beta_m.
\end{equation*}
The variation of \eqref{5.3} with respect to $l$ and $\beta_m$ yield the two additional boundary conditions at $x=l$
\begin{equation}
\begin{split}\label{5.5}
\beta_1^\prime(l)=0,
\\ 
2f^\prime(0)-(\gamma_l(\gamma)-\beta_m)(c-2l)=0.
\end{split}
\end{equation}
The first condition of \eqref{5.5} means the continuity of the dislocation density.
It becomes clear from this construction that the above variational problem has no solution for the unmodified energy density $f(y)$ from \eqref{3.1}. For the modified function $f(y)$ from \eqref{3.2} the variational problem is well-posed and there exist a unique minimizer. 

It is obvious that $l\to 0$ when $\beta _m\to 0$. In this limit we can find the critical value $\gamma _c$, at which the excess dislocations begin to pile up, as the root of the equation
\begin{equation}
\label{5.10}
\gamma -g(\gamma )=2(\phi^*-\frac{1}{4\pi}\ln k_0)/c.
\end{equation}
This equation shows clearly the size effect. For $\gamma >\gamma_c$, the system \eqref{5.4}-\eqref{5.5} has non-trivial solution. Since the integrand in functional \eqref{5.3} does not depend on $x$, Euler's equation \eqref{5.4} admits the first integral
\begin{displaymath}
\frac{1}{2}(\gamma_l-\beta_1)^2+f(\beta _1^\prime )-\beta _1^\prime f^\prime (\beta _1^\prime )=C.
\end{displaymath}
With $f(y)$ from \eqref{3.3} and with the boundary conditions $\beta_1^\prime (l)=0$ and $\beta _1(l)=\beta_m$, this equation reduces to
\begin{equation}\label{5.10a}
\frac{(\beta_1^{\prime })^2}{4\pi } \left( k_1-\frac{2}{k_0+\beta _1^\prime }\right) =(\gamma_l-\beta_1)^2-(\gamma_l-\beta_m)^2.
\end{equation}
Due to the convexity of $f(\beta ^\prime )$ the left-hand side of \eqref{5.10a} is a monotonously increasing function of $\beta _1^\prime $. Therefore, for each $\beta _1<\beta_m$ there exist a unique root $\beta _1^\prime $ of this equation. The numerical solution of \eqref{5.10a} will be discussed in Section 6. After finding the average plastic slip we calculate the average shear stress according to
\begin{align}
\bar{\tau}&=\frac{1}{c}\int_0^c \mu (\gamma -\beta(x))\dd{x} \notag
\\
&=\frac{\mu}{c}\left[ \gamma c -2\int_0^l \beta _1(x) \dd{x}-\beta_m (c-2l)\right]. \label{5.11}
\end{align}

\section{Numerical simulations}

\begin{figure}[htb]
	\centering
	\includegraphics[width=7cm]{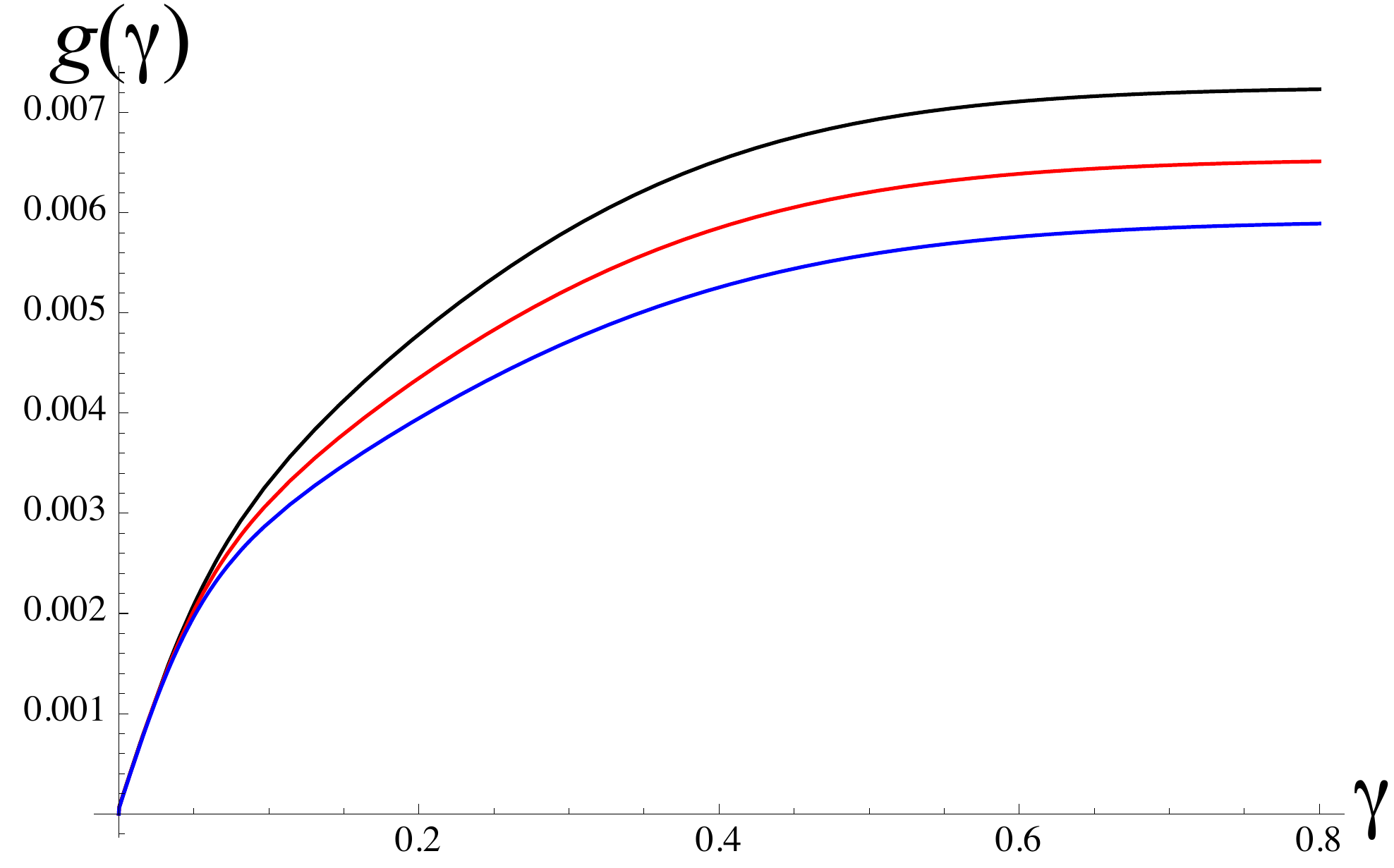}
	\caption{Functions $g(\gamma )=\tau_Y(\gamma)/\mu$: (i) $\tilde{q}_0=10^{-12}$ (black),  (ii) $\tilde{q}_0=10^{-14}$ (red), (i) $\tilde{q}_0=10^{-16}$ (blue).}
	\label{fig:3}
\end{figure}

Assume that the crystal is loaded with the constant shear strain rate $\dot{\gamma }$. As discussed in the previous Section, the first task is then to solve the system \eqref{eq:4.11}, with $\tau $ being replaced by $\tau _Y$ and $q_0=t_0 \dot{\gamma }$. Since the shear strain rate $\dot{\gamma }$ is constant and only $g(\gamma )=\tau_Y(\gamma)/\mu$ is required for the next task, we choose $\gamma $ as the independent variables and rewrite this system of equations in terms of the following dimensionless quantities
\begin{equation*}
g(\gamma )=\frac{\tau _Y(\gamma )}{\mu },\quad \tilde{\rho }=a^2\rho , \quad \tilde{\chi }=\frac{\chi }{e_D}, \quad \tilde{\rho }^{ss}(\tilde{\chi})=e^{-1/\tilde{\chi}}.
\end{equation*}

\begin{figure}[htb]
	\centering
	\includegraphics[width=7cm]{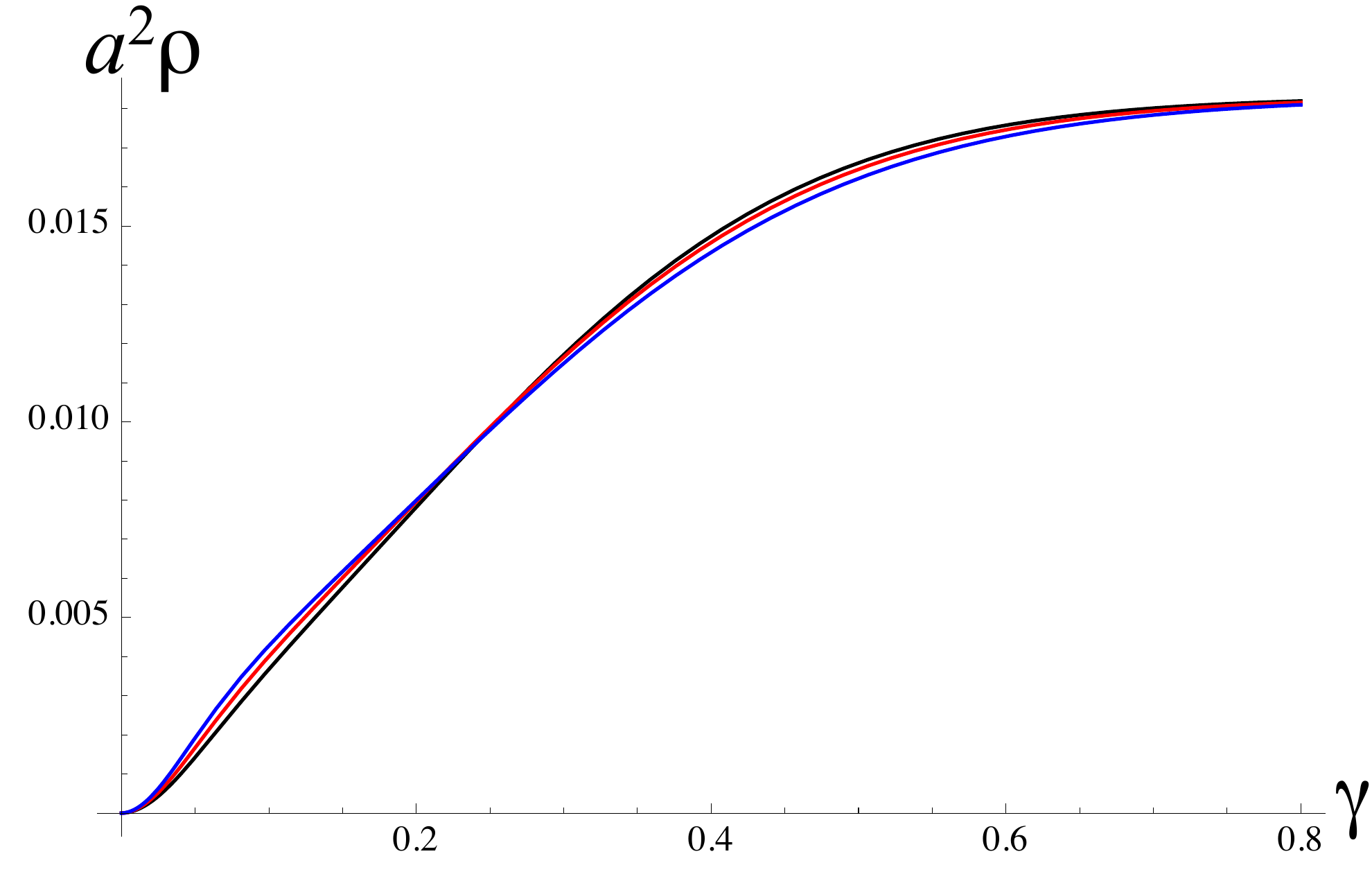}
	\caption{Normalized total density of dislocations $\tilde{\rho }(\gamma )=a^2\rho $: (i) $\tilde{q}_0=10^{-12}$ (black),  (ii) $\tilde{q}_0=10^{-14}$ (red), (i) $\tilde{q}_0=10^{-16}$ (blue).}
	\label{fig:3a}
\end{figure}

The system of ODEs becomes
\begin{align}
\frac{\mathrm{d}g}{\mathrm{d}\gamma } & = 1-\frac{\tilde{q}(g,\tilde{\rho })}{\tilde{q}_0} ,  \notag
\\
\frac{\mathrm{d}\tilde{\chi }}{\mathrm{d}\gamma } & = K g \frac{\tilde{q}(g,\tilde{\rho })}{\tilde{q}_0} \left[ 1-\frac{\tilde{\chi }}{\tilde{\chi }^{ss}(\tilde{q})} \right], \label{eq:6.2}
\\
\frac{\mathrm{d}\tilde{\rho }}{\mathrm{d}\gamma } & = \frac{K_\rho g }{\tilde{\nu }(\theta ,\tilde{\rho },\tilde{q}_0)^2}\frac{\tilde{q}(g,\tilde{\rho })}{\tilde{q}_0}\left[ 1-\frac{\tilde{\rho }}{\tilde{\rho }^{ss}(\tilde{\chi })} \right] . \notag
\end{align}

\begin{figure}[htb]
	\centering
	\includegraphics[width=7cm]{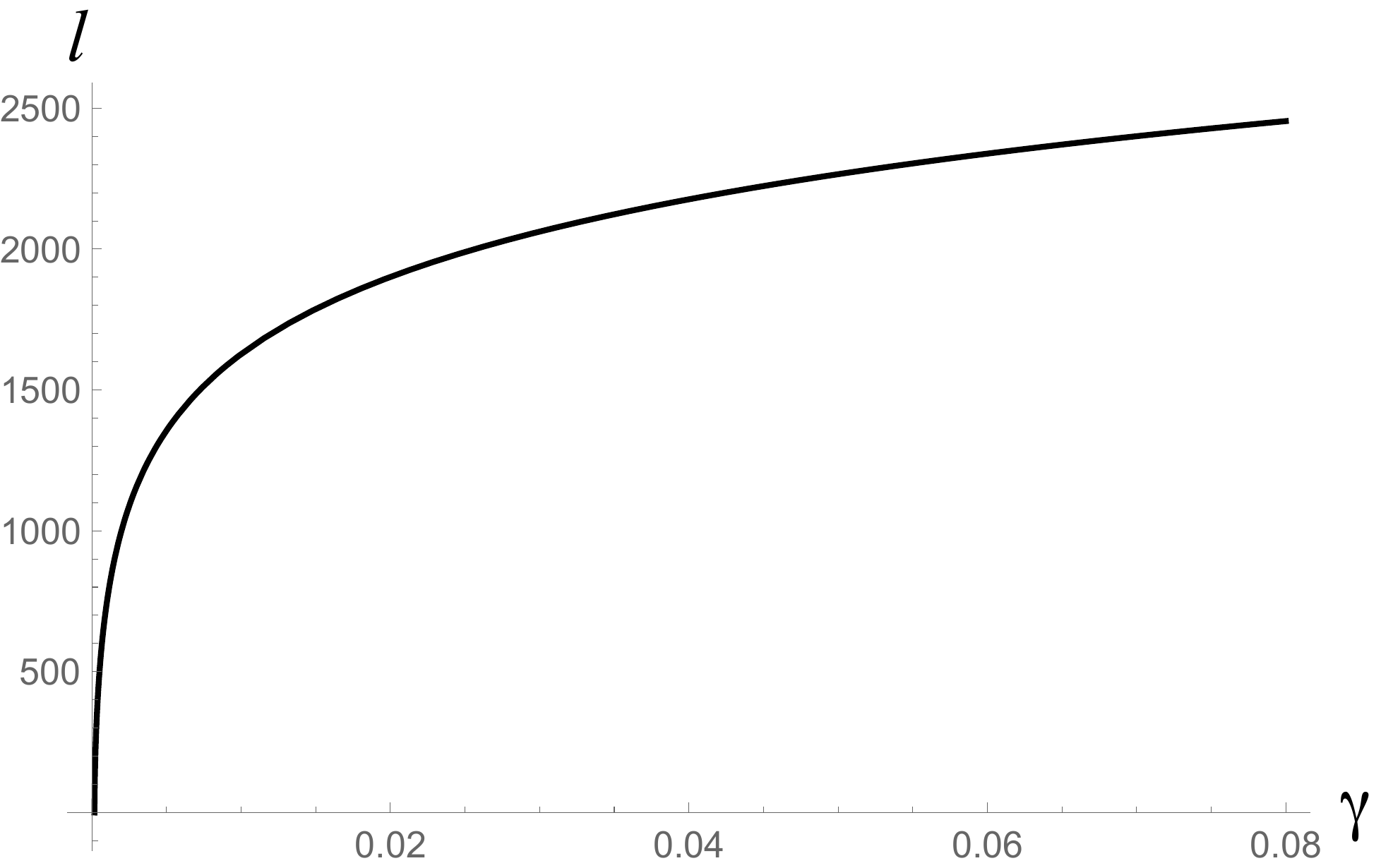}
	\caption{Relative thickness of the boundary layers $l(\gamma )$.}
	\label{fig:4}
\end{figure}

Here $\tilde{q}_0=(a/b)\dot{\gamma }t_0$, $r=(b/a)\mu_T/\mu $, $K=\mathcal{K}\mu $, $K_\rho =\mathcal{K}_\rho \mu $ and
\begin{align*}
\tilde{q}(g,\tilde{\rho }) & = \sqrt{\tilde{\rho }}\exp \left[ -\frac{1}{\theta }e^{-g/(r\sqrt{\tilde{\rho }})}\right] , \\
\tilde{\nu }(\theta ,\tilde{\rho },\tilde{q}_0) & = \ln \left( \frac{1}{\theta }\right) - \ln \left[ \frac{1}{2}\ln \left( \frac{\tilde{\rho }}{\tilde{q}_0^2}\right) \right]  .
\end{align*}
Let $T=298$ K. The parameters for copper at this room temperature are chosen as follows \cite{Langer2010} 
\begin{equation*}
r=0.0323, \quad \theta =0.0073, \quad K=350, \quad K_\rho =96.1,\quad \tilde{\chi }=0.25.
\end{equation*}
We choose also the initial conditions 
\begin{equation*}
g(0)=0,\quad \tilde{\rho}(0)=10^{-6}, \quad \tilde{\chi}(0)=0.18.
\end{equation*}
The plots of functions $g(\gamma )$ found by the numerical integration of \eqref{eq:6.2} for three different resolved shear strain rates are shown in Fig.~\ref{fig:3}. It can be seen that $g(\gamma )$ is rate-sensitive. Besides, it is also temperature-sensitive. Fig.~\ref{fig:3a} shows the evolution of the normalized total density of dislocations $a^2\rho $ versus $\gamma $ for the above shear strain rates.

\begin{figure}[htb]
	\centering
	\includegraphics[width=7cm]{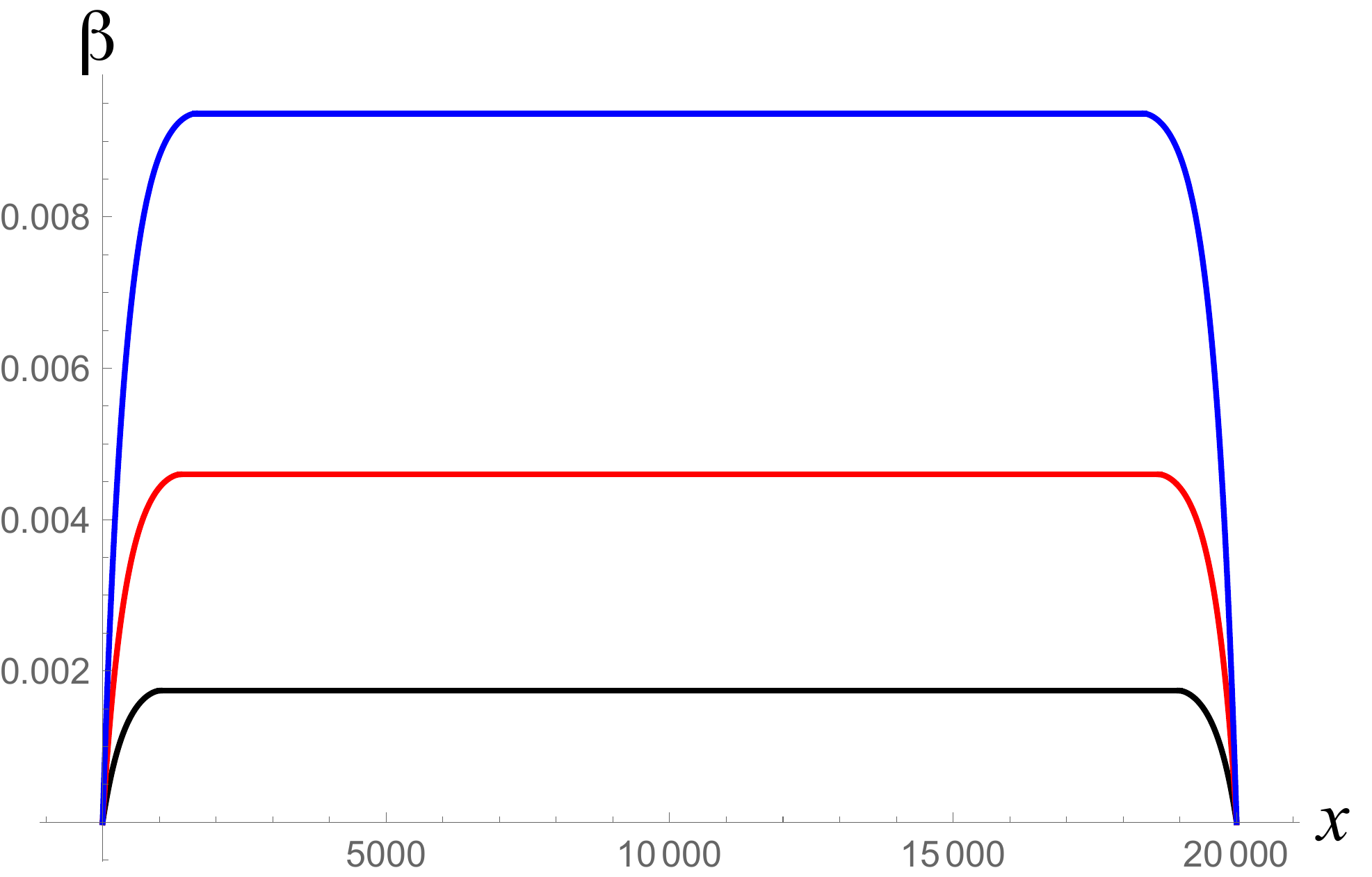}
	\caption{The plastic slip $\beta(x)$: (i) $\gamma =0.002$ (black), (ii) $\gamma =0.005$ (red), (iii) $\gamma =0.01$ (blue).}
	\label{fig:5}
\end{figure}

Having found $g(\gamma )$, we turn now to the determination of the plastic slip $\beta (x)$ from the energy minimization problem \eqref{5.2}. In this problem let us fix $\tilde{q}_0=10^{-12}$ and choose the following parameters for copper
\begin{equation*}
b=0.255\, \text{nm}, \quad a=2.55\, \text{nm},\quad c=5.1\, \mu\text{m},\quad \nu=0.355.
\end{equation*}
We also choose $k_0=10^{-6}$, $k_1=2.1\times 10^6$. With these parameters and with the above function $g(\gamma )$ we find from equation \eqref{5.10} that $\gamma _c=0.00016$. To solve equation \eqref{5.10a} we reduce it to the following cubic equation
\begin{displaymath}
k_1 q^3+(k_0k_1-2)q^2-4\pi \alpha q-4\pi k_0 \alpha =0,
\end{displaymath}
with $q=\beta_1^{\prime }$ and $\alpha =(\gamma_l-\beta_1)^2-(\gamma_l-\beta_m)^2$. Due to the convexity of $f(\beta_1^{\prime })$ for the chosen set of parameters, this cubic equation has only one positive real root that we denote by
\begin{displaymath}
\beta_1^{\prime }=p(\alpha )=p((\gamma_l-\beta_1)^2-(\gamma_l-\beta_m)^2).
\end{displaymath}
Integrating this equation numerically, we find 
\begin{equation}
\label{6.6}
x=\int_0^{\beta _1}\frac{\dd{z}}{p((\gamma_l-z)^2-(\gamma_l-\beta_m)^2)}
\end{equation}
which is the inverse function of $\beta _1(x)$. The length of the boundary layer equals
\begin{equation}
\label{6.7}
l(\beta _m)=\int_0^{\beta _m}\frac{\dd{z}}{p((\gamma_l-z)^2-(\gamma_l-\beta_m)^2)}.
\end{equation}

\begin{figure}[htb]
	\centering
	\includegraphics[width=7cm]{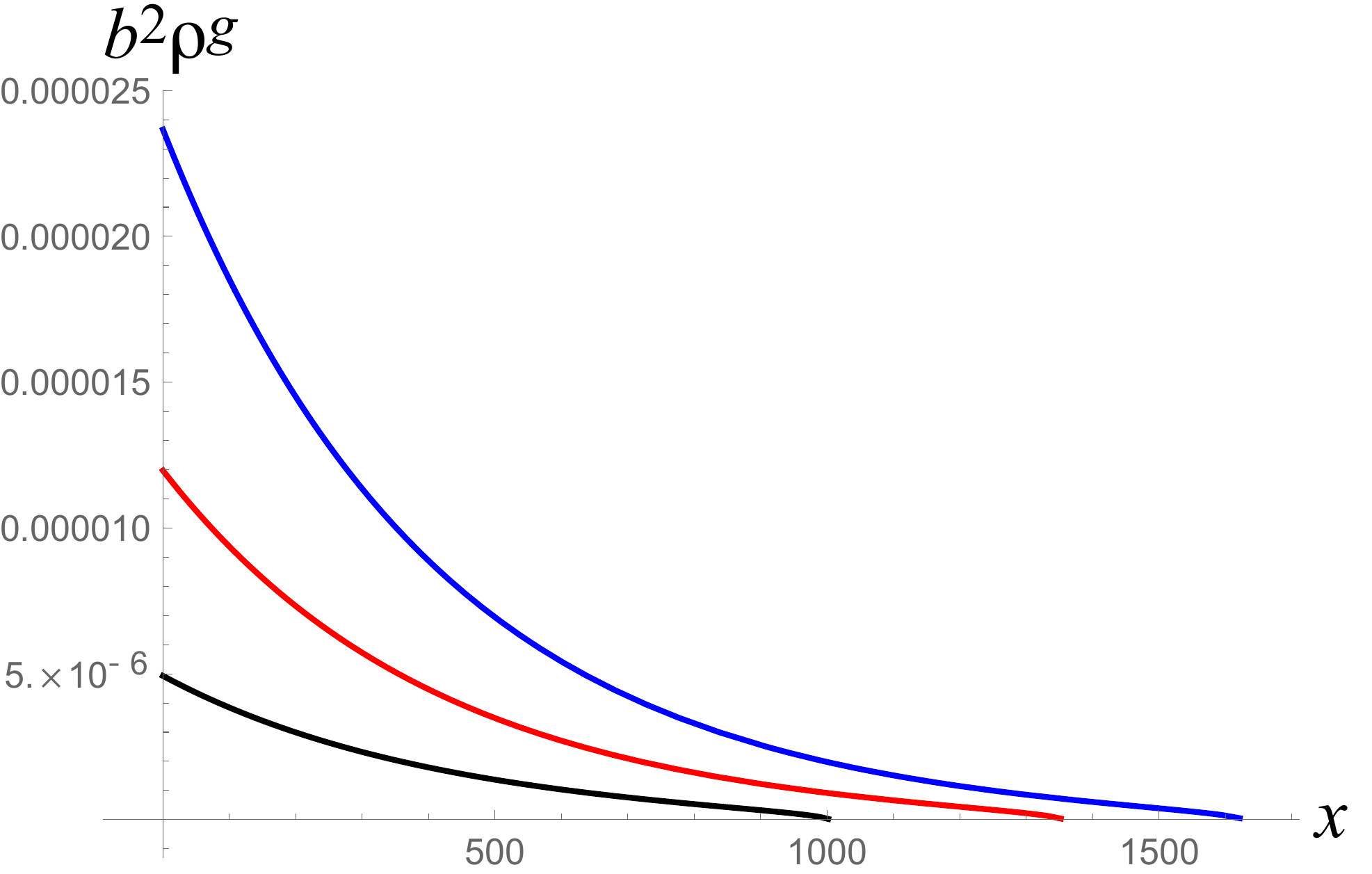}
	\caption{The normalized density of excess dislocations $b^2\rho ^g=\beta ^\prime $: (i) $\gamma =0.002$ (black), (ii) $\gamma =0.005$ (red), (iii) $\gamma =0.01$ (blue).}
	\label{fig:5a}
\end{figure}

\begin{figure}[htb]
	\centering
	\includegraphics[width=7cm]{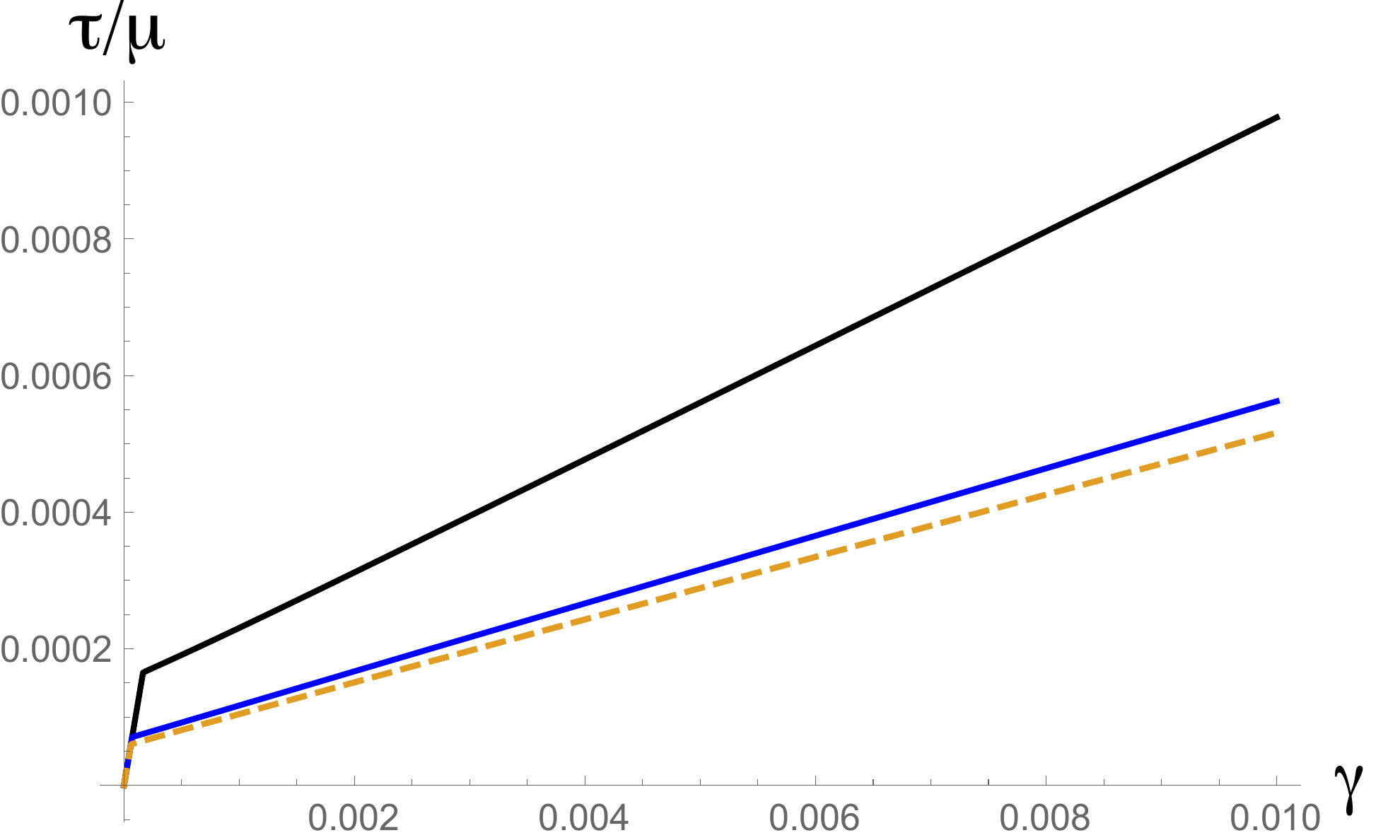}
	\caption{Dimensionless average shear stress versus shear strain curve: (i) present theory: $c=5.1$ micron (black), $c=51$ micron (blue) (ii) LBL-theory (dashed).}
	\label{fig:6}
\end{figure}

Substituting function $l(\beta _m)$ from \eqref{6.7} into \eqref{5.5}$_2$, we get an equation to determine $\beta _m$ in terms of $\gamma $. Then, with this $\beta _m(\gamma)$ we find also the length $l(\gamma)=l(\beta_m(\gamma))$ of the boundary layer. The plot of function $l(\gamma)$ is shown in Fig.~\ref{fig:4}, from which it is seen that $l(\gamma )$ is a monotonically increasing function of $\gamma $. However, for the whole range of $\gamma \in (\gamma _{c},0.08)$ the relative thickness of the boundary layers $l$ remains small compared to $\bar{c}=2\times 10^4$. Next, we find with \eqref{6.6} the plastic slip as function of $x$ at three chosen values of $\gamma >\gamma _{c}$. Their plots are shown in Fig.~\ref{fig:5}. We see that the plastic slip is constant in the middle of the specimen, and changes rapidly only in two thin boundary layers where the positive and negative excess dislocations pile up against the grain boundaries. The number of excess dislocations increases with increasing shear strain. It is interesting to know the distribution of normalized density of excess dislocations $b^2\rho ^g=\beta _1^\prime $. Their distributions at three chosen values of $\gamma >\gamma _{c}$ and in the left boundary layer are shown in Fig.~\ref{fig:5a}. In the right boundary layer the excess dislocations of opposite sign are symmetrically distributed. Using the implicit equation \eqref{6.6} we reduce equation \eqref{5.11} for the average dimensionless shear stress to
\begin{equation*}
\tau/\mu =\gamma -\frac{1}{c}\left[ 2\int_0^{\beta _m} \frac{\beta \dd{\beta }}{p((\gamma_l-\beta)^2-(\gamma_l-\beta_m)^2)}+\beta_m(c-2l)\right] .
\end{equation*}
The dimensionless shear stress versus the shear strain curve computed in accordance with this equation is shown in Fig.~\ref{fig:6} for two chosen widths of the sample: (i) $c=5.1$ micron (black), (ii) $c=51$ micron (blue). For comparison we show also the stress-strain curve $g(\gamma )$ computed in accordance with the LBL-theory (dashed line). We see that, in addition to the isotropic work-hardening caused by the redundant dislocations, there is a kinematic work-hardening caused by the pile-up of excess dislocations against the grain boundaries. The difference due to this kinematic work-hardening becomes remarkable at large strains. Besides, it is seen that this kinematic work-hardening decreases when the thickness of the specimen increases, thus exhibiting the size effect (cf. \cite{Berdichevsky-Le07}). For single crystals of macroscopic sizes the kinematic work-hardening is negligibly small, and the stress-strain curve approaches that of LBL-theory.

\section{Conclusion}
In this paper we develop the thermodynamic dislocation theory for non-uniform plastic deformations of crystals undergoing anti-plane constrained shear. The asymptotically exact energy density found by the averaging procedure is extrapolated to the extremely small or large dislocation density. In the problem of anti-plane constrained shear, the excess dislocations are concentrated in thin boundary layers near the grain boundaries. The stress-strain curves exhibit both the isotropic hardening due to the redundant dislocations and kinematic hardening due to the pile-ups of excess dislocations against the grain boundaries which is size-dependent. For single crystals of macroscopic sizes the kinematic work-hardening is negligibly small, and the stress-strain curve approaches that of LBL-theory.

\end{document}